\documentclass[a4paper,11pt,twoside]{article}
\usepackage[british]{babel}
\usepackage{amssymb,amsmath,amsfonts,verbatim,xkeyval,bm}
\usepackage{graphicx}
\usepackage{subfigure}
\usepackage{textcomp}
\newcommand{\pluseq}{\hookleftarrow}
\def\vet#1{\mathbf#1}
\def\build#1_#2^#3{\mathrel{
\mathop{\kern 0pt#1}\limits_{#2}^{#3}}}
\def\imunit{{\rm i}}
\newcommand{\csi}{\xi}
\newcommand{\Cscr}{{\cal C}}
\newcommand{\Gscr}{{\cal G}}
\newcommand{\Hscr}{{\cal H}}
\newcommand{\Kscr}{{\cal K}}
\newcommand{\Lscr}{{\cal L}}

\newcommand{\Pscr}{{\cal P}}
\newcommand{\Rscr}{{\cal R}}

\newcommand{\Zscr}{{\cal Z}}
\def\poisson#1#2{\lbrace #1,#2 \rbrace}
\def\Lie{{\Lscr}}

\newtheorem{theorem}{Theorem}[section]

\newtheorem{proposition}[theorem]{Proposition}

\title{\bf Design of maneuvers based on new normal form
  approximations: the case study of the CPRTBP\thanks{ {\it Key words
      and phrases:} Restricted three-body problem, normal forms,
    Hamiltonian perturbation theory, averaging, Celestial Mechanics
    averaging, impulsive transfer, maneuvers design}}

\author{{\bf Roc\'io Isabel P\'AEZ}\\ {\small Dipartimento di
    Matematica, Universit\`a degli Studi di Roma ``Tor
    Vergata'',}\\ {\small Via della Ricerca Scientifica 1, 00133--Roma
    (Italy).}\\ {\bf Ugo LOCATELLI}\\ {\small Dipartimento di
    Matematica, Universit\`a degli Studi di Roma ``Tor
    Vergata'',}\\ 
  {\small Via della Ricerca Scientifica 1, 00133--Roma
    (Italy).}\\ 
}

\date{}

\begin{document}
\maketitle

\bigskip

\markboth{R.I. P\'aez, U. Locatelli}{Maneuvers based on integrable approximation}

\begin{abstract}
In this work, we study the motions in the region around the
equilateral Lagrangian equilibrium points $L_4$ and $L_5$, in the
framework of the Circular Planar Restricted Three-Body Problem
(hereafter, CPRTBP). We design a semi-analytic approach based on some
ideas by Garfinkel in~\cite{Garfinkel-77}: the Hamiltonian is expanded
in Poincar\'{e}--Delaunay coordinates and a suitable average is
performed.  This allows us to construct (quasi) invariant tori that
are moderately far from the Lagrangian points L4-L5 and approximate
wide tadpole orbits. This construction provides the tools for studying
optimal transfers in the neighborhood of the equilateral points, when
instantaneous impulses are considered. We show some applications of
the new averaged Hamiltonian for the Earth-Moon system, applied to the
setting-up of some transfers which allow to enter in the stability
region filled by tadpole orbits.
\end{abstract}

\maketitle


\section{Introduction}\label{sec:intro}

For the design of simple transfers in Astrodynamics, Hohmann transfers
are widely used.  They consist basically in a maneuvre in a system
represented by a Two--Body Problem (2BP) and their solutions,
consequently given by Keplerian ellipses. Starting from a circular
orbit around a main body, a transfer to a different (inner or outer)
circular orbit can be achieved with just two different impulses of
properly defined sense and magnitude (see e.g. \cite{Perozzi-2003}).

In cases where 2BP is not a suitable approximation, a similar approach
can be considered with a different simplified version of the
Hamiltonian representing the system. The aim of the method lays on the
idea of designing a transfer between orbits that are {\it exact
  solutions} of an {\it integrable approximation} of the studied
system, using a set of impulses. Since, in general, physical problems
have more than one degree of freedom, a suitable construction of a
normal form approximating the Hamiltonian of the system is
needed. Furthermore, many d.o.f. make representations in configuration
space inadequate, since they do not give precise hints of the time
evolution of the orbits at glance. Thus, the baseline is the
construction of a normalized integrable approximation of the model to
study, that should provide: i)~analytical solutions for the motions,
and ii)~suitable surfaces of sections, tools to be used for the
analysis of the effects of the impulses that conform the trasfer.\\

In the last decades, several semi-analytical results have been
obtained in order to ensure the stability of the motions of some Trojan
asteroids, orbiting around the Lagrangian equilibrium points $L_4-L_5$
of the Sun--Jupiter system. All those works share a same common
structure, summarized as follows: each approach is based
on an explicit algorithm, that can be translated on a computer so as to
calculate the expansion of a suitable normal form, providing a good
local approximation of the complete Hamiltonian of the CPRTBP. Such a
normal form is used to approximate the orbits of some objects and to
prove their stability, provided they are close enough to the
equilateral Lagrangian points. As far as we know, the wider coverage
of the tadpole\footnote{For an introduction to tadpole and horseshoe
  orbits in the CPRTBP model, see, e.g., \S~3.9
  of~\cite{MurrDermm-1999}.} orbits around $L_4-L_5$ is given
in~\cite{Gab-Jor-Loc-05}, that is based on the Kolmogorov normal
form. Here, we try to improve those results, by revisiting the
approach developed in~\cite{Garfinkel-77}. From that article we borrow
two main ideas: first, we perform the initial expansions of the CPRTBP
Hamiltonian in a suitable set of Poincar\'{e}--Delaunay--like
coordinates (while polar coordinates were used in~\cite{Gab-Jor-Loc-05}). 
That particular type of canonical variables allows us to
clearly distinguish a pair of slowly varying coordinates from those
quickly changing their values. Therefore, we average the Hamiltonian
with respect to the angle related to the fast dynamics.  This second
main idea is implemented here, by using the modern Lie series
formalism so as to construct an {\it integrable} normal form; this allows
us to approximate also the tadpole orbits going rather far from
$L_4-L_5\,$. In fact, as a major novelty with respect
to~\cite{Garfinkel-77} (that is based on purely analytical
techniques), we have translated our procedure in some codes, producing
suitably truncated expansions of the normal form and, therefore,
explicit numerical results.

Having the tools described before, we create an algorithm which is
able to test the effectiveness of different impulses, after choosing
an arbitrary starting point.  This approach can be applied in many
different systems. In particular, we focus our work in the area
surrounding the Lagrangian equilibrium points $L_4-L_5$ of the
Earth-Moon system. This kind of stability region is interesting since
it is located close to the Earth, and could provide a natural
trapping zone for small bodies as astronomical observatories, obsolet
spacecrafts or space debris in general.  At the end of the paper, we
characterize the effects of different instantaneous impulses and we
choose the best candidates for effective transfers in the framework
provided by the normal form approximating the CPRTBP Hamiltonian.

\section{Explicit construction of the integrable approximation}\label{sec:expl_alg}

\subsection{Initial settings}\label{sbs:settings}

Let us introduce the standard framework of the CPRTBP, as done, e.g.,
in~\cite{Efthy-2013}. In such a model, the motion of the two biggest
bodies (hereafter, the primaries) is not influenced by the third one,
considered massless, so the orbits of the primaries are Keplerian
ellipses and the third body moves under the gravitational attraction
exerted by the other two. Additionally, we assume that both those
Keplerian orbits are circular and the massless body is coplanar with
the primaries. Let us define the heliocentric vectors
$\vet{r}_{j0}=\vet{x}_{j}-\vet{x}_{0}$ with $j=1\,,\,2\,$, being
$\vet{x}_{0}\,$, $\vet{x}_{1}$ and $\vet{x}_{2}$ the position vectors
of the biggest primary, the smallest one and the third body,
respectively, in an {\em inertial} frame.  As usual, we also define
the units of measures in order to set the rotation period $T=2\pi$ and
the gravitational constant $\Gscr=1$, and we denote with $\mu$ and
$1-\mu$ the masses of the smallest primary and the largest one,
respectively.  These settings imply that the semi-major axis of the
ellipse described by the orbit of $\vet{r}_{10}$ is equal to $1$,
while $\mu\in(0,1/2)$.

For our purposes, it is convenient to
adopt the Hamiltonian formalism in the non-inertial {\em synodic}
frame that co-rotates with the primaries. We define the axes so
that the largest primary is always located at the origin of the
synodic frame, while the fixed position of the smallest primary is
such that $\vet{r}_{10}=(-1,0)$. Let us first introduce the
action--angle canonical coordinates $(G,\Gamma,\lambda,\ell)$, which can
be seen as modified Delaunay variables for the {\em planar Keplerian
  problem}. They are given by
\begin{equation}
\vcenter{\openup1\jot\halign{
 \hbox {\hfil $\displaystyle {#}$}
&\hbox {\hfil $\displaystyle {#}$\hfil}
&\hbox {$\displaystyle {#}$\hfil}
&\hbox {\hfil $\displaystyle {#}$}
&\hbox {\hfil $\displaystyle {#}$\hfil}
&\hbox {$\displaystyle {#}$\hfil}\cr
G &=& \sqrt{a(1-e^2)}\ ,
\qquad
&\lambda &=& M+g-M^{\prime}-g^{\prime}\ ,
\cr
\Gamma &=& \sqrt{a}\left(1-\sqrt{1-e^2}\right)\ ,
\qquad
&\ell &=& M\ ,
\cr
}}
\label{eq:Delaunay-coord}
\end{equation}
where $a\,$, $e\,$, $M$ and $g$ are the semi-major axis, the
eccentricity, the mean anomaly and the longitude of the perihelion of
the massless body, being the angle $g$ measured in the {\it inertial}
frame. Moreover, $M^{\prime}$ and $g^{\prime}$ denote the mean anomaly
and the perihelion longitude of the smallest primary,
respectively. Thus, $\lambda$ corresponds to the {\it synodic mean
  longitude}.  The Hamiltonian ruling the motion of the third body in
the non-inertial synodical frame can now be written as follows:
\begin{equation}
\Hscr(G,\Gamma,\lambda,\ell)=
-\frac{1}{2(G+\Gamma)^2}-G
-\mu F(G,\Gamma,\lambda,\ell)\ ,
\label{eq:initial-synod-Ham}
\end{equation}
where the so--called disturbing function $\mu F$ is such that
\begin{equation}
F=\frac{1}{\|\vet{r}_{20}\|}-\frac{1}{\|\vet{r}_{20}-\vet{r}_{10}\|}
+\frac{\vet{r}_{10}\cdot\vet{r}_{20}}{\|\vet{r}_{10}\|^3}\ .
\label{eq:perturbing-term}
\end{equation}

In order to remove the singularity of the Delaunay--like
variables when $e=0$ (i.e., for circular orbits), it is
convenient to introduce canonical coordinates
$(\rho,\csi,\lambda,\eta)$,  similar to the Poincar\'e coordinates
for the {\em planar Keplerian problem}. Thus, let us define
\begin{equation}
\vcenter{\openup1\jot\halign{ \hbox {\hfil $\displaystyle {#}$} &\hbox
    {\hfil $\displaystyle {#}$\hfil} &\hbox {$\displaystyle {#}$\hfil}
    &\hbox {\hfil $\displaystyle {#}$} &\hbox {\hfil $\displaystyle
      {#}$\hfil} &\hbox {$\displaystyle {#}$\hfil}\cr \rho &= & G-1\ ,
    \qquad &\lambda &= & \lambda\ , \cr \csi &=
    &\sqrt{2\Gamma}\cos\ell\ , \qquad &\eta &=
    &\sqrt{2\Gamma}\sin\ell\ , \cr }}
\label{eq:Garfinkel-coord}
\end{equation}
where the values of $\rho$ are significantly small in a region
surrounding the Lagrangian points, for instance, in the case of
tadpole  or horseshoe orbits. In fact, in those cases, it holds true since
$a\simeq 1$ and $e\gtrsim 0$.  Let us recall that those variables
have been adopted in~\cite{Garfinkel-77} to study the CPRTBP.

Before constructing a normal form, the starting
Hamiltonian~\eqref{eq:initial-synod-Ham} (in particular, the
disturbing function~\eqref{eq:perturbing-term}) must be expanded in
Poincar\'e--Delaunay--like coordinates $(\rho,\csi,\lambda,\eta)$.
Such a non trivial operation is described in~\cite{Paez-2013}, which
also includes all the detailed {\tt Mathematica} codes explicitly
producing those expansions. This approach ensures that the starting
Hamiltonian $\Hscr$ can be written in the following form as a function
of the Poincar\'e--Delaunay--like coordinates:

\begin{equation}
H^{(0,0)}(\rho,\csi,\lambda,\eta)=
\sum_{l\ge 0}Z_l^{(0)}\big(\rho,(\csi^2+\eta^2)/2\big)
\,+\,\sum_{s\ge 1}\sum_{l\ge 0}\mu^sf_l^{(0,0;s)}(\rho,\csi,\lambda,\eta)\ ,
\label{eq:H(0,0)}
\end{equation}
where $Z_l^{(0)}\in\Pscr_{l,0}$ and $f_l^{(0,0;s)}\in\Pscr_{l,sK}$
$\forall\ l\ge 0,\ s\ge 1$, being $K$ a fixed positive integer.
$\Pscr_{l,sK}$ is the set of functions such that
\begin{itemize}
\item a function $g\in\Pscr_{l,sK}$ if the generic terms appearing in
  its Taylor--Fourier expansion, which are of type
  $c_{m_1,m_2,m_3,k}\rho^{m_1}\csi^{m_2}\eta^{m_3}\cos(k\lambda)$ or
  $d_{m_1,m_2,m_3,k}\rho^{m_1}\csi^{m_2}\eta^{m_3}\sin(k\lambda)$,
  satisfy the following relations about their coefficients:
$$
c_{m_1,m_2,m_3,k}=d_{m_1,m_2,m_3,k}=0
\qquad{\rm when}\qquad
2m_1+m_2+m_3\neq l
\quad{\rm or}\quad
|k|>sK\ .
$$
\end{itemize}
\noindent
In principle, the expansion~\eqref{eq:H(0,0)} can be seen as a
reorganization of the Taylor--Fourier series giving the disturbing
function; this is made in a suitable way that allows us to
successfully perform the construction of the normal form. Furthermore,
the criterion for the choice of $K$ is such that the size (in any
common functional norm) of $f_l^{(0,0;1)}$, $f_l^{(0,0;2)}$, $\ldots$
is approximately the same for any value of the index $l$. In other
words, the disturbing function is splitted in terms $O(\mu)$,
$O(\mu^2)$, $\ldots$ by using the Fourier decay of the coefficients
for increasing values of the harmonic $|k|$; let us recall that in the
present work $\mu$ is regarded as a fixed small parameter of the
system.

\subsection{Averaging the Hamiltonian over the fast angle}\label{sbs:average}

Let us emphasize that the expansion~\eqref{eq:H(0,0)} contains also a
non-trivial information about the Keplerian part (that corresponds to
the whole Hamiltonian if $\mu=0$). In fact, since
$Z_l^{(0)}=Z_l^{(0)}\big(\rho,(\csi^2+\eta^2)/2\big)$ and
$Z_l^{(0)}\in\Pscr_{l,0}\,$, then one can deduce that $Z_l^{(0)}=0$
when the index $l$ is odd. This is in agreement with the expansion of
the starting Hamiltonian $\Hscr$, when the disturbing function is
neglected and the actions $G$ and $\Gamma$ are substituted according
to formula~\eqref{eq:Garfinkel-coord}. If we explicitely write the
first main terms of the Keplerian part
\begin{equation}
Z_0^{(0)}+Z_2^{(0)}+Z_4^{(0)}=
-\frac{3}{2}+\frac{\csi^2+\eta^2}{2}
-\frac{3}{2}\left[\rho+\frac{\csi^2+\eta^2}{2}\right]^2=
-\frac{3}{2}+\Gamma-\frac{3}{2}(\rho+\Gamma)^2\ ,
\label{eq:quadr-part-Kepl-appr}
\end{equation}
we conclude that the angular velocities have different
order of magnitude
\begin{equation}
\dot\lambda=\frac{\partial\,\Hscr}{\partial\rho}\simeq 0\ ,
\qquad
\dot\ell=\frac{\partial\,\Hscr}{\partial\Gamma}\simeq 1\ ,
\label{eq:slow-fast-ang-velocities}
\end{equation}
because $\mu\ll 1$ and the values of the actions $\rho$ and $\Gamma$
are small in a region surrounding the Lagrangian points. Thus,
$\lambda$ can be seen as a slow angle and $\ell$ as a fast
angle. Then, this motivates to average the Hamiltonian over the fast
angle (see, e.g., \S~52 of~\cite{Arnold-book-analyt-mech}), in order
to focus mainly on the secular evolution of the system. We remove all
the terms depending on the fast angle $\ell$, by performing a sequence
of canonical transformations. In the following, this strategy will be
translated in an explicit algorithm. Such a procedure will allow us to
produce a final Hamiltonian satisfying two important properties: at
the same time it provides a good approximation of the starting system,
it only depens on the actions and one of the angles, i.e., it is
integrable.

\subsubsection{Construction of the averaged normal form: the formal algorithm}\label{sss:formal-algorithm}

As discussed above, the normalization algorithm defines a
sequence of Hamiltonians. This is done by an iterative procedure; let
us describe the basic step which introduces $H^{(r_1,r_2)}$ starting
from $H^{(r_1,r_2-1)}$ when both the values of the indexes $r_1$ and
$r_2$ are positive. We assume that the expansions of $H^{(r_1,r_2-1)}$
is such that
\begin{equation}
\vcenter{\openup1\jot\halign{
 \hbox {\hfil $\displaystyle {#}$}
&\hbox {\hfil $\displaystyle {#}$\hfil}
&\hbox {$\displaystyle {#}$\hfil}\cr
H^{(r_1,r_2-1)}(\rho,\csi,\lambda,\eta) &=
&\sum_{s=0}^{r_1-1}\sum_{l\ge 0}
  \mu^s Z_l^{(s)}\big(\rho,(\csi^2+\eta^2)/2,\lambda\big)
\,+\,\sum_{l=0}^{r_2-1}\mu^{r_1}
 Z_l^{(r_1)}\big(\rho,(\csi^2+\eta^2)/2,\lambda\big)
\cr
& &+\sum_{l\ge r_2}\mu^{r_1}f_l^{(r_1,r_2-1;r_1)}(\rho,\csi,\lambda,\eta)
\,+\,\sum_{s>r_1}\sum_{l\ge 0}\mu^sf_l^{(r_1,r_2-1;s)}(\rho,\csi,\lambda,\eta)\ ,
\cr
}}
\label{eq:H(r1,r2-1)}
\end{equation}
where $Z_l^{(s)}\in\Pscr_{l,sK}$ $\forall\ l\ge 0,\ 0\le s<r_1\,$,
$Z_l^{(r_1)}\in\Pscr_{l,r_1K}$ $\forall\ 0\le l< r_2\,$,
$f_l^{(r_1,r_2-1;r_1)}\in\Pscr_{l,r_1K}$ $\forall\ l\ge r_2\,$,
$f_l^{(r_1,r_2-1;s)}\in\Pscr_{l,sK}$ $\forall\ l\ge 0,\ s>r_1\,$.  Let
us recall that the Hamiltonian~\eqref{eq:H(0,0)} is suitable to start
the procedure with $r_1=r_2=1$, after having set
$H^{(1,0)}=H^{(0,0)}$.  In formula~\eqref{eq:H(r1,r2-1)}, one can
 distinguish the normal form terms from the perturbing part; the
latter depends on $(\csi,\eta)$ in a {\it generic} way, while in the
first two terms of~\eqref{eq:H(r1,r2-1)} the fast
variables can be replaced by the action
$\Gamma=(\csi^2+\eta^2)/2\,$. The $(r_1,r_2)$--th step of the
algorithm formally defines the new Hamiltonian as
\begin{equation}
H^{(r_1,r_2)}=\exp\Big(\Lie_{\mu^{r_1}\chi_{r_2}^{(r_1)}}\Big)H^{(r_1,r_2-1)}\ ,
\label{eq:def-funzionale-H(r1,r2)}
\end{equation}
where $\exp\big(\Lie_{\chi}\big)\,\cdot=\sum_{j\ge
  0}\frac{1}{j!}\Lie_{\chi}^j\,\cdot$ is the Lie series operator, with
$\Lie_{\chi} g=\poisson{g}{\chi}$ (being $\poisson{\cdot}{\cdot}$ the
classical Poisson bracket), $g$ a generic function defined on the
phase space and $\chi$ any generating function (for an introduction to
canonical transformations by Lie series in the context of the
Hamiltonian perturbation theory, see,
e.g.,~\cite{Giorgilli-2003.1}). The new generating function
$\mu^{r_1}\chi_{r_2}^{(r_1)}(\rho,\csi,\lambda,\eta)$ is determined so
as to remove from the main perturbing term\footnote{Let us recall that
  the size of $\mu^{s}f_{r_2}^{(r_1,r_2-1;s)}\in\Pscr_{l,sK}$ is
  expected to decrease when the indexes $s$ or $r_2$ are increased,
  because the values of $\mu$, $\rho$ and $\sqrt{\csi^2+\eta^2}$ are
  assumed to be small} $\mu^{r_1}f_{r_2}^{(r_1,r_2-1;r_1)}$ its
subpart that is not in normal form. This is done by solving the
following homological equation with respect to
$\chi_{r_2}^{(r_1)}=\chi_{r_2}^{(r_1)}(\rho,\csi,\lambda,\eta)$:
\begin{equation}
\Lie_{\chi_{r_2}^{(r_1)}}Z_2^{(0)}+f_{r_2}^{(r_1,r_2-1;r_1)}=Z_{r_2}^{(r_1)}\ ,
\label{eq:chi(r1,r2-1)}
\end{equation}
where we require that $Z_{r_2}^{(r_1)}$ is the new term in normal
form, i.e.
$Z_{r_2}^{(r_1)}=Z_{r_2}^{(r_1)}\big(\rho,(\csi^2+\eta^2)/2,\lambda\big)$.

\begin{proposition}
When $Z_2^{(0)}=(\csi^2+\eta^2)/2$ and
$f_{r_2}^{(r_1,r_2-1;r_1)}\in\Pscr_{r_2,r_1K}\,$, then there exists a
generating function $\chi_{r_2}^{(r_1)}\in\Pscr_{r_2,r_1K}$ and a
normal form term $Z_{r_2}^{(r_1)}\in\Pscr_{r_2,r_1K}$ solving the
homological equation~\eqref{eq:chi(r1,r2-1)}.
\label{lem:sol_homol_eq}
\end{proposition}
We limit ourselves to just sketch the procedure that can be followed
so as to explicitly determine a solution of~\eqref{eq:chi(r1,r2-1)}
and, therefore, prove the statement above. First, we replace the fast
coordinates $(\csi,\eta)$ with the pair of complex conjugate canonical
variables $(z,\imunit{\overline z})$ such that $\csi=(z-{\overline
  z})/\sqrt{2}$ and $\eta=(z+{\overline z})/\sqrt{2}$. Moreover, the
homological equation~\eqref{eq:chi(r1,r2-1)} has to be expanded in
Taylor series with respect to $(z,\imunit{\overline z})$, using the
slow coordinates $(\rho,\lambda)$ as fixed parameters (because they
are not affected by the Poisson bracket
$\Lie_{\chi_{r_2}^{(r_1)}}Z_2^{(0)}$, since $Z_2^{(0)}$ do not depend
on them). Therefore, we solve term-by-term the
equation~\eqref{eq:chi(r1,r2-1)} in the unknown coefficients
$x_{m_2,m_3}(\rho,\lambda)$ and $\zeta_{m}(\rho,\lambda)$ such that
$$
\chi_{r_2}^{(r_1)}(\rho,z,\lambda,\imunit{\overline z})=
\sum_{m_2,m_3}\big[x_{m_2,m_3}(\rho,\lambda)z^{m_2}(\imunit{\overline z})^{m_3}\big]
\ ,\qquad
Z_{r_2}^{(r_1)}(\rho,z,\lambda,\imunit{\overline z})=
\sum_{m}\big[\zeta_{m}(\rho,\lambda)z^{m}(\imunit{\overline z})^{m}\big]\ .
$$
At last, we express the expansions above by replacing $(z,\imunit{\overline z})$ for  $(\csi,\eta)$,
so as to obtain the final solutions in the form
$\chi_{r_2}^{(r_1)}=\chi_{r_2}^{(r_1)}(\rho,\csi,\lambda,\eta)$ and
$Z_{r_2}^{(r_1)}=Z_{r_2}^{(r_1)}\big(\rho,(\csi^2+\eta^2)/2,\lambda\big)$.

The following property of the Poisson brackets is very useful
for our purposes, and since the proof is immediate, it is omitted.

\begin{proposition}
Let $f$ and $g$ be two generic functions such that $f\in\Pscr_{r,sK}$
and $g\in\Pscr_{r^{\prime},s^{\prime}K}\,$, then
$$
{\rm if}\ r+r^{\prime}\ge 2\ \ \Rightarrow
\ \ \poisson{f}{g}\in\Pscr_{r+r^{\prime},(s+s^{\prime})K}\ ,
\qquad
{\rm else}\ \ \Rightarrow\ \ \poisson{f}{g}=0\ .
$$
\label{lem:alg-prop-Pois-brackets}
\end{proposition}

\noindent 
In order to provide an algorithm easy to translate in a
programming language, we are going to give explicit formulas for the
new Hamiltonian $H^{(r_1,r_2)}$ and for its expansion that can be written
as follows:
\begin{equation}
\vcenter{\openup1\jot\halign{
 \hbox {\hfil $\displaystyle {#}$}
&\hbox {\hfil $\displaystyle {#}$\hfil}
&\hbox {$\displaystyle {#}$\hfil}\cr
H^{(r_1,r_2)}(\rho,\csi,\lambda,\eta) &=
&\sum_{s=0}^{r_1-1}\sum_{l\ge 0}
  \mu^s Z_l^{(s)}\big(\rho,(\csi^2+\eta^2)/2,\lambda\big)
\,+\,\sum_{l=0}^{r_2}\mu^{r_1}
 Z_l^{(r_1)}\big(\rho,(\csi^2+\eta^2)/2,\lambda\big)
\cr
& &+\sum_{l\ge r_2+1}\mu^{r_1}f_l^{(r_1,r_2;r_1)}(\rho,\csi,\lambda,\eta)
\,+\,\sum_{s>r_1}\sum_{l\ge 0}\mu^sf_l^{(r_1,r_2;s)}(\rho,\csi,\lambda,\eta)\ ,
\cr
}}
\label{eq:H(r1,r2)}
\end{equation}
For the sake of simplicity in the calculation of $f_l^{(r,r_2;s)}$, we
redefine the same quantity several times using the same symbol.  This
notation of the algorithm is more similar to its translation in a
programming code, and, thus, more useful: let us introduce the
recursive operation $a\pluseq b$, where the previously defined
quantity $a$ is redefined as $a=a+b\,$. Therefore, we initially define
\begin{equation}
f_l^{(r_1,r_2;s)}=f_l^{(r_1,r_2-1;s)}
\qquad\ \forall\ l>r_2\ {\rm when}\ s=r_1
\ \ \ {\tt or}\ \ \ \forall\ l\ge 0\,,\ s\ge r_1\,.
\label{eq:f_l^r1r2s_def_1}
\end{equation}
Then, we consider the contribution of the terms generated by the Lie
series applied to each function belonging to the normal form part as
follows:
\begin{equation}
f_{l+j(r_2-2)}^{(r_1,r_2;s+jr_1)}\pluseq
\frac{1}{j!}\Lie_{\chi_{r_2}^{(r_1)}}^jZ_l^{(s)}
\qquad\ \forall\ 1\le j< {\bar j}_f\,,
\ 0\le l< {\bar l}_f\,,\ 0\le s\le r_1\ ,
\label{eq:f_l^r1r2s_def_2}
\end{equation}
where the upper limits ${\bar j}_f$ and ${\bar l}_f$ on the indexes
$j$ and $l$, respectively, are such that
\begin{equation}
\vcenter{\openup1\jot\halign{
 \hbox {\hfil $\displaystyle {#}$}
&\hbox {$\displaystyle {#}$\hfil}
\qquad
&\hbox {\hfil $\displaystyle {#}$}
&\hbox {$\displaystyle {#}$\hfil}
\cr
{\bar j}_f=l+1 &\ {\rm if}\ r_2=1\ ,
&{\bar j}_f=+\infty &\ {\rm if}\ r_2\ge 2\ ,\cr
{\bar l}_f=+\infty &\ {\rm if}\ s<r_1\ ,
&{\bar l}_f=r_2 &\ {\rm if}\ s=r_1\ ,\cr
{\bar l}_i=r_2 &\ {\rm if}\ s=r_1\ ,
&{\bar l}_i=0 &\ {\rm if}\ s>r_1\ .\cr
}}
\label{eq:limits-indexes-j-l}
\end{equation}
For what concerns the contributions given by the perturbing terms
making part of the expansion of $H^{(r_1,r_2-1)}$
in~\eqref{eq:H(r1,r2-1)}, we have
\begin{equation}
f_{l+j(r_2-2)}^{(r_1,r_2;s+jr_1)}\pluseq
\frac{1}{j!}\Lie_{\chi_{r_2}^{(r_1)}}^jf_l^{(r_1,r_2-1;s)}
\qquad\ \forall\ 1\le j< {\bar j}_f\,,
\ l\ge {\bar l}_i\,,\ s\ge r_1\ ,
\label{eq:f_l^r1r2s_def_3}
\end{equation}
where the limiting values for the indexes, that are ${\bar j}_f$ and
${\bar l}_i\,$, are defined in~\eqref{eq:limits-indexes-j-l}.

The redefinition rules~\eqref{eq:f_l^r1r2s_def_2}
and~\eqref{eq:f_l^r1r2s_def_3} are set so that the new perturbing part generated
by the Lie series in~\eqref{eq:def-funzionale-H(r1,r2)} is coherently
split in different terms according to their order of magnitude in
$\mu$ and their total polynomial degree in the actions. In fact, by
applying repeatedly proposition~\ref{lem:alg-prop-Pois-brackets} to
the redefinitions
in~\eqref{eq:f_l^r1r2s_def_1}--\eqref{eq:f_l^r1r2s_def_3}, it is
possible to inductively verify that
$f_l^{(r_1,r_2;s)}\in\Pscr_{l,sK}$ $\forall\ l\ge {\bar l}_i,\ s\ge
r_1\,$. Therefore, the terms making part of the
Hamiltonian~$H^{(r_1,r_2)}$ in the expansion~\eqref{eq:H(r1,r2)} share
the same properties with those appearing in~\eqref{eq:H(r1,r2-1)};
this ensures that the normalization algorithm can be iterated so as to
construct $H^{(r_1,r_2+1)},\ H^{(r_1,r_2+2)},\ \ldots$

\subsubsection{Criteria for stopping the normalization algorithm, in order to perform a finite number of operations}\label{sss:stop-algorithm}

From an ideal point of view, we would be interested in producing the
final Hamiltonian
$\lim_{r_1\to\infty}\lim_{r_2\to\infty}H^{(r_1,r_2)}$, where
$H^{(r_1+1,0)}$ is defined as $\lim_{r_2\to\infty}H^{(r_1,r_2)}$
$\forall\ r_1\ge 1$. In fact, such a Hamiltonian would be integrable,
because it would depend from the fast variables just through the
action $(\csi^2+\eta^2)/2$, due to the special form of the
expansion~\eqref{eq:H(r1,r2)}. In general, the problem\footnote{We
  emphasize that the most celebrated problem concerning the
  convergence of the normal forms, the accumulation of
  ``small divisors'', does not affect our scheme, because the main
  integrable term in the homological equation~\eqref{eq:chi(r1,r2-1)},
  i.e. $Z_2^{(0)}$, depends just on the fast action
  $(\csi^2+\eta^2)/2$.} of the restrictions of domains prevents the
convergence of the limits (with respect to both indexes $r_1$ and
$r_2\,$, when the standard $\sup$--norm is used); thus, it is not
possible to define the integrable Hamiltonian on any open set (see,
e.g.,~\cite{Giorgilli-2003.1}). Actually, from a mathematical point of
view, expansions of type~\eqref{eq:H(r1,r2)} are {\it asympotic
  series} with respect to both $r_1$ and $r_2\,$; this means that we
have to truncate the indexes, but in a way that optimize our
result. For this purpose, we can proceed as follows. First, we
introduce the functions $\Zscr^{(r_1,r_2)}
=\Zscr^{(r_1,r_2)}\big(\rho,(\csi^2+\eta^2)/2,\lambda\big)$ and
$\Rscr^{(r_1,r_2)}=\Rscr^{(r_1,r_2)}(\rho,\csi,\lambda,\eta)$ that
make explicit the splitting between the integrable and the perturbing
parts in the expansion~\eqref{eq:H(r1,r2)}, so that
\begin{equation}
\Zscr^{(r_1,r_2)} = \sum_{s=0}^{r_1-1}\sum_{l\ge 0} \mu^s Z_l^{(s)}
\,+\,\sum_{l=0}^{r_2}\mu^{r_1} Z_l^{(r_1)}\ ,
\qquad
\Rscr^{(r_1,r_2)} = \sum_{l\ge r_2+1}\mu^{r_1}f_l^{(r_1,r_2;r_1)}
\,+\,\sum_{s>r_1}\sum_{l\ge 0}\mu^sf_l^{(r_1,r_2;s)}\ .
\label{eq:split-integr-and-perturb-parts-in-H(r1,r2)}
\end{equation}
Therefore, we look for the pair of upper indexes $(R_1,R_2)$
minimizing the $\sup$--norm of $\Rscr^{(r_1,r_2)}$ (on the set of
values of the variables that we are interested to study). This
approach can be implemented with a suitable scheme of analytic
estimates, in order to reduce exponentially the remainder
$\Rscr^{(R_1,R_2)}$ with respect to the small parameters of the
problem (see~\cite{Nekhoroshev-1977}, \cite{Nekhoroshev-1979} and
~\cite{Giorgilli-2003.1}). Such an optimal choice about the final
values of the indexes $r_1$ and $r_2$ allows us to reformulate the
algorithm, in such a way that it requires just $R_1R_2$ normalization
steps, constructing the finite sequence of Hamiltonians
$H^{(0,0)}=H^{(1,0)},\ H^{(1,1)},\ \ldots\,,\ H^{(1,R_2)},\ \ldots\,,
\ H^{(R_1,0)},\ H^{(R_1,1)},\ \ldots\,,\ H^{(R_2,R_1)}$, where
$H^{(r_1+1,0)}=H^{(r_1,R_2)}$ $\forall\ 1\le r_1<R_1\,$.

While the algorithm has been rearranged so as to be performed in a
finite number of normalization steps, it is evident that the
redefinition rules reported in formulas~\eqref{eq:f_l^r1r2s_def_2}
and~\eqref{eq:f_l^r1r2s_def_3} would require to calculate infinitely
many Poisson brackets. In order to avoid such a problem, we have to
establish two truncation rules on the terms appearing in the
expansions, so as to fix (a)~their maximal exponent $s_{max}$ related
to the order of magnitude $O(\mu^s)$,~(b) their total maximal degree
$l_{max}$ on the index $l$ (that is equal to twice the degree in
$\rho$ plus the one in $\csi$ and that in $\eta$). If our formal
algorithm is subject to a further restriction, such that it is limited
to the calculation of functions belonging to classes of type
$\Pscr_{l,sK}$ with $0\le l\le l_{max}$ and $0\le s\le s_{max}\,$,
then it can be proved that it requires a {\it finite} total number of
Poisson brackets\footnote{Each of them needs a finite number of basic
  operations like derivatives, sums and products.}. Therefore, this
newly restricted version of our algorithm is suitable to be translated
in a programming code. In principle, the values of $l_{max}$ and
$s_{max}$ should be chosen in order to optimize the final results; in
practice, they are usually fixed (as well as the final indexes $R_1$
and $R_2$) so as to fit with the available computational resources.

\subsubsection{Approximate numerical integration based on the normalizing canonical transformation}\label{sss:semi-analytical_integr-scheme}

It is well known that Lie series induce canonical transformations in
a Hamiltonian framework; this fundamental feature will allow us to
design a numerical integration method, by using both the normal form
 discussed above and the corresponding canonical
coordinates. In order to explicitly realize such a project, we have to
introduce some more complicate notations. Let us denote with
$\big(\rho^{(r_1,r_2)},\csi^{(r_1,r_2)},\lambda^{(r_1,r_2)},\eta^{(r_1,r_2)}\big)$
the set of canonical coordinates related to the $(r_1,r_2)$--th step.
By appying the so called exchange theorem (see,
e.g.,~\cite{Giorgilli-2003.1}), we have that
\begin{equation}
H^{(r_1,r_2)}
\big(\rho^{(r_1,r_2)},\csi^{(r_1,r_2)},\lambda^{(r_1,r_2)},\eta^{(r_1,r_2)}\big)
=H^{(r_1,r_2-1)}\Big(\varphi^{(r_1,r_2)}
\big(\rho^{(r_1,r_2)},\csi^{(r_1,r_2)},\lambda^{(r_1,r_2)},\eta^{(r_1,r_2)}\big)\Big)\ ,
\label{eq:exchange-theorem}
\end{equation}
where the variables related to the previous step, namely
$\big(\rho^{(r_1,r_2-1)},\csi^{(r_1,r_2-1)},\lambda^{(r_1,r_2-1)},\eta^{(r_1,r_2-1)}\big)$,
are given as
\begin{equation}
\varphi^{(r_1,r_2)}
\big(\rho^{(r_1,r_2)},\csi^{(r_1,r_2)},\lambda^{(r_1,r_2)},\eta^{(r_1,r_2)}\big)=
\exp\Big(\Lie_{\mu^{r_1}\chi_{r_2}^{(r_1)}}\Big)
\big(\rho^{(r_1,r_2)},\csi^{(r_1,r_2)},\lambda^{(r_1,r_2)},\eta^{(r_1,r_2)}\big)\ ;
\label{eq:coord-change-(r1,r2)}
\end{equation}
the r.h.s. of the equation above means that four Lie series
must be applied separatedly to each variable, in order to properly
define all the coordinates for the canonical transformation
$\varphi^{(r_1,r_2)}$. The whole normalization procedure can be
described by the canonical transformation
\begin{equation}
\Cscr^{(R_2,R_1)}=\varphi^{(1,1)}\circ\ldots\circ\varphi^{(1,R_2)}\circ\ldots
\circ\varphi^{(R_1,1)}\ldots\circ\varphi^{(R_2,R_1)} \ .
\label{eq:def-total-canonical-transf}
\end{equation}
Such a composition of all the intermediate changes of variables can be
used for providing the following semi-analytical scheme to integrate
the equations of motion:
\begin{equation}
\vcenter{\openup1\jot\halign{
  \hbox to 36 ex{\hfil $\displaystyle {#}$\hfil}
&\hbox to 12 ex{\hfil $\displaystyle {#}$\hfil}
&\hbox to 36 ex{\hfil $\displaystyle {#}$\hfil}\cr
\left(\rho^{(0,0)}(0),\csi^{(0,0)}(0),
\lambda^{(0,0)}(0),\eta^{(0,0)}(0)\right)
&\build{\longrightarrow}_{}^{{{\scriptstyle
\big(\Cscr^{(R_1,R_2)}\big)^{-1}}
\atop \phantom{0}}}
&\left(\rho^{(R_1,R_2)}(0),\csi^{(R_1,R_2)}(0),
\lambda^{(R_1,R_2)}(0),\eta^{(R_1,R_2)}(0)\right)
\cr
& &\big\downarrow \build{\Phi_{\Zscr^{(R_1,R_2)}}^{t}}_{}^{}
\cr
\left(\rho^{(0,0)}(t),\csi^{(0,0)}(t),
\lambda^{(0,0)}(t),\eta^{(0,0)}(t)\right)
&\build{\longleftarrow}_{}^{{{\scriptstyle \Cscr^{(R_1,R_2)}} \atop \phantom{0}}}
&\left(\rho^{(R_1,R_2)}(t),\csi^{(R_1,R_2)}(t),
\lambda^{(R_1,R_2)}(t),\eta^{(R_1,R_2)}(t)\right)
\cr
}}
\qquad\qquad\ ,
\label{semi-analytical_scheme}
\end{equation}
where $\Phi_{\Kscr}^{t}$ is the flow induced on the canonical
coordinates by the generic Hamiltonian $\Kscr$ for an interval of time
equal to $t$. Let us emphasize that the above integration scheme
provides an approximate solution; from an ideal point of view (i.e.,
if all the expansions were performed without errors and truncations),
formula~\eqref{semi-analytical_scheme} would be exact if
$\Zscr^{(R_1,R_2)}$ would correspond to the complete Hamiltonian
$H^{(R_1,R_2)}$. On the other hand, $\Zscr^{(R_1,R_2)}$ is {\it
  integrable} and its flow is easy to compute\footnote{In order to
  explicitly describe the solutions of the equation of motions for the
  normal form $\Zscr^{(R_1,R_2)}$, it is convenient to introduce the
  temporary action--angle variables
  $\big(\Gamma^{(R_1,R_2)},\ell^{(R_1,R_2)}\big)$ such that
  $\csi^{(R_1,R_2)}=\sqrt{2\Gamma^{(R_1,R_2)}}\cos\ell^{(R_1,R_2)}$
  and
  $\eta^{(R_1,R_2)}=\sqrt{2\Gamma^{(R_1,R_2)}}\sin\ell^{(R_1,R_2)}$, where $\Gamma^{(R_1,R_2)}$
  is a constant of motion for the normal form
  $\Zscr^{(R_1,R_2)}=\Zscr^{(R_1,R_2)}\big(\rho^{(R_1,R_2)},\Gamma^{(R_1,R_2)},\lambda^{(R_1,R_2)}\big)$. By
  considering $\Gamma^{(R_1,R_2)}$ as a fixed parameter and using the
  standard quadrature method for conservative systems with $1$~d.o.f.,
  one can compute $\rho^{(R_1,R_2)}(t)$ and
  $\lambda^{(R_1,R_2)}(t)$ at any time $t\,$. The same can be done for
  the evolution of $\ell^{(R_1,R_2)}(t)$, by evaluating the integral
  corresponding to the differential equation
  ${\dot\ell}^{(R_1,R_2)}=\frac{\partial\,\Zscr^{(R_1,R_2)}}{\partial\Gamma^{(R_1,R_2)}}\,$. For
  practical purposes, the application of the classical quadrature
  method can be replaced by any numerical integrator that is precise
  enough. Finally, the values of $\csi^{(R_1,R_2)}(t)$ and
  $\eta^{(R_1,R_2)}(t)$ can be directly calculated from those of the
  corresponding action--angle variables, that are
  $\Gamma^{(R_1,R_2)}(t)$ and $\ell^{(R_1,R_2)}(t)$.}, reasons why
using $\Zscr^{(R_1,R_2)}$ becomes valuable.  The approximate solution
provided by the scheme~\eqref{semi-analytical_scheme} is as more
accurate as smaller the perturbing part $\Rscr^{(R_1,R_2)}$ is with
respect to $\Zscr^{(R_1,R_2)}$ (see their definitions
in~\eqref{eq:split-integr-and-perturb-parts-in-H(r1,r2)}).

As a final remark, we stress that it requires a {\it finite} total
number of operations, if we limit the the expansions of the canonical
transformations as discussed in the previous subsection. Thus, also
the whole integration scheme~\eqref{semi-analytical_scheme} can be
translated in a programming code.

\subsection{Tests on the accuracy of the normal form approximating the CPRTBP Hamiltonian}\label{sbs:results_sect}

In order to test the accuracy of our integrable normal form
approximating the Hamiltonian, we have to choose a suitable surface of
section for the comparison with the complete problem.  The most
logical choice about the sectioning of the flow is given by the
surface defined by $\ell = 0$, because $\ell$ is a fast angle (recall
the discussion about formula~\eqref{eq:slow-fast-ang-velocities}).

\begin{figure}
  \includegraphics[height=.3\textheight,angle=270]{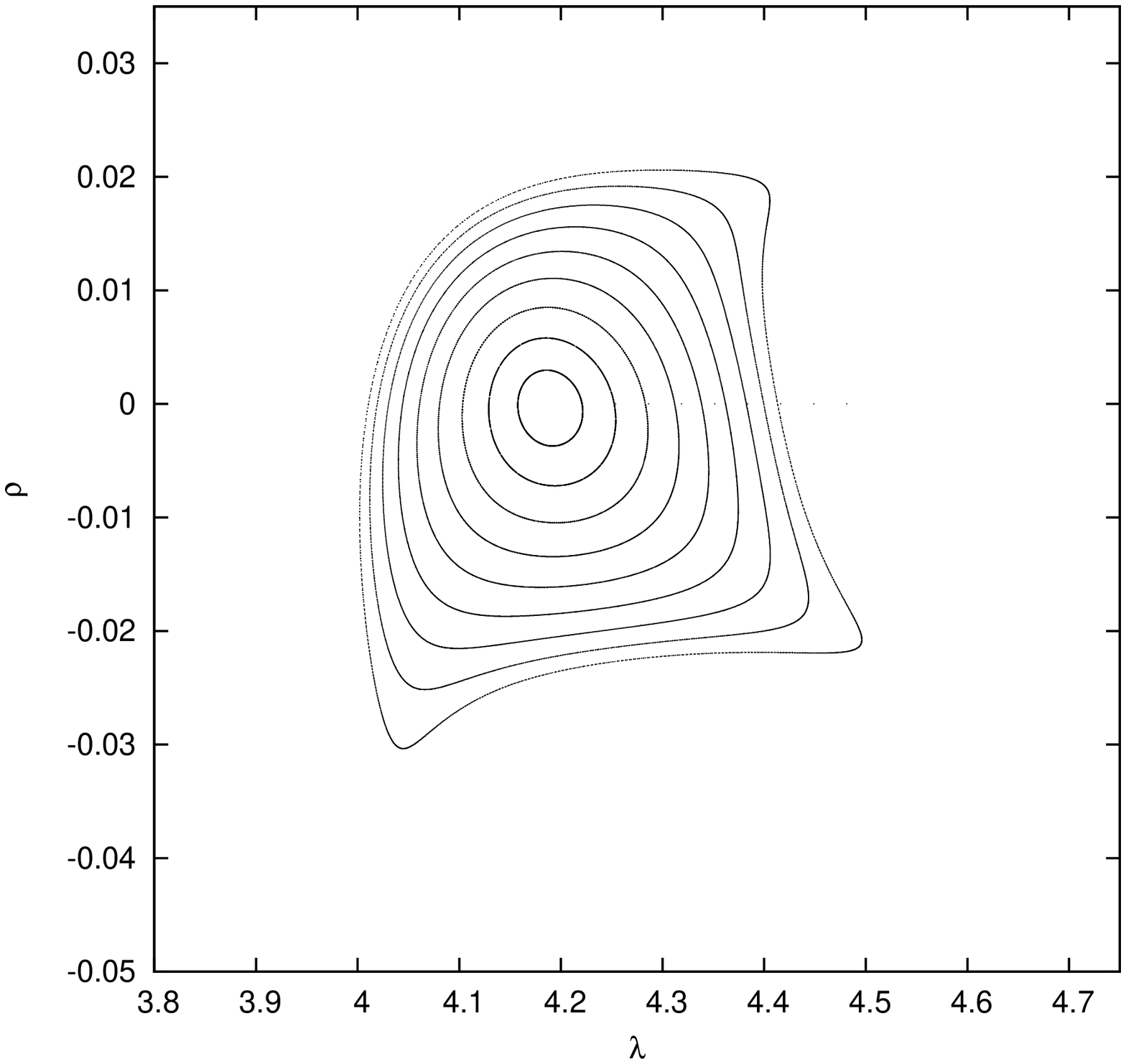}
  \hspace{-1cm}
  \includegraphics[height=.3\textheight,angle=270]{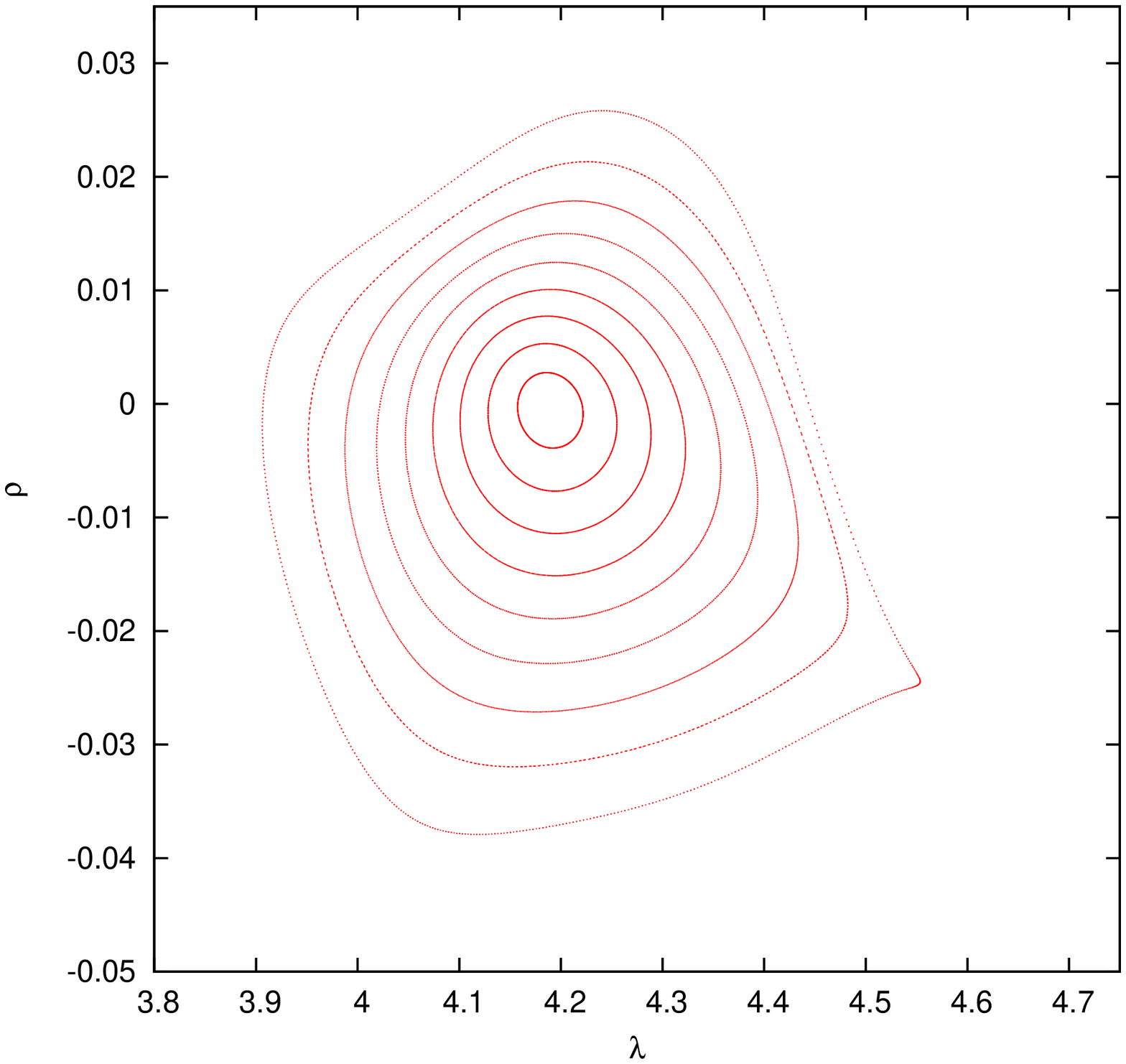}
  \hspace{-1cm}
  \includegraphics[height=.3\textheight,angle=270]{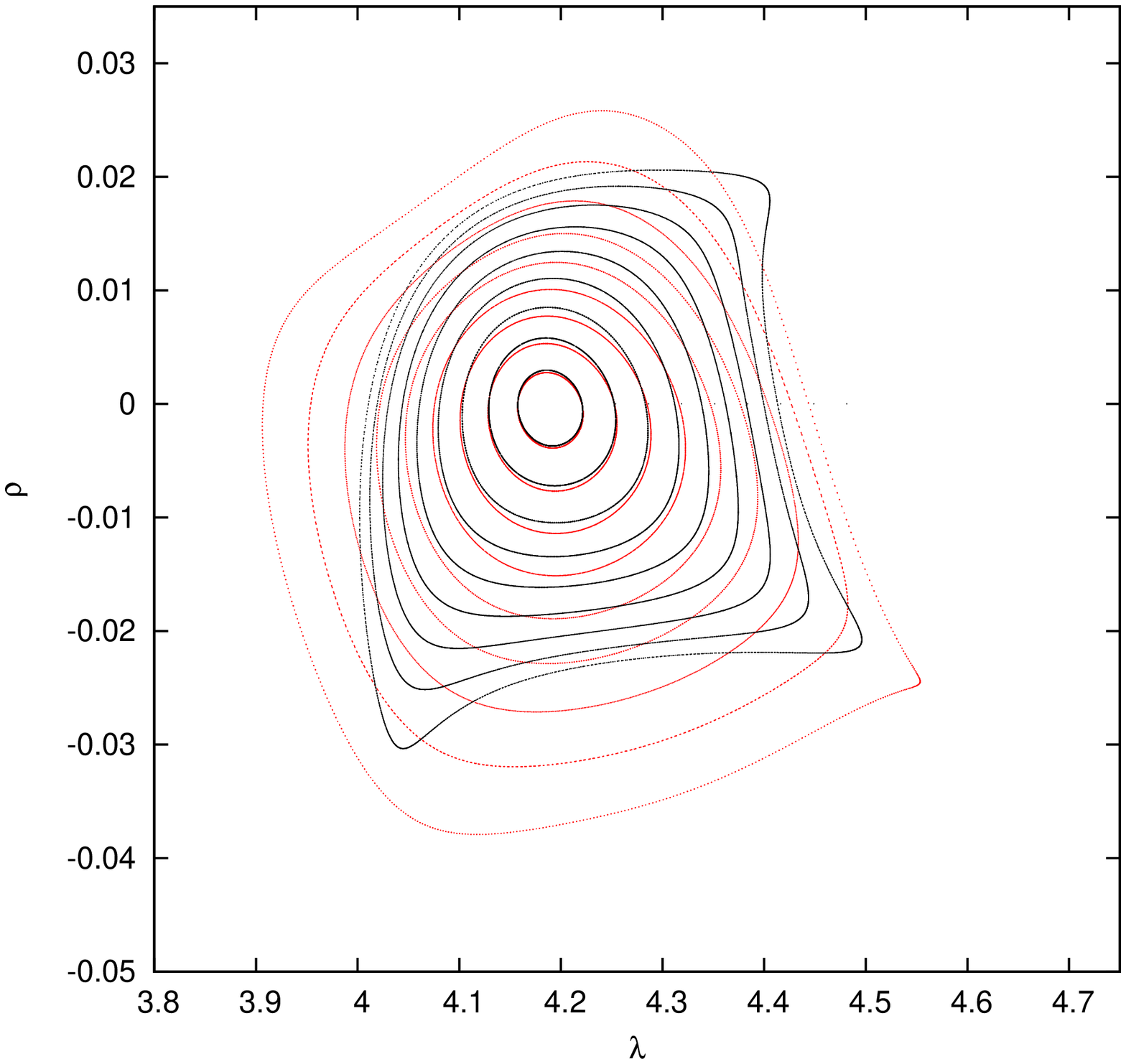}
  \caption{Left panel, in black surface of section obtained by a
    numerical integration of the CPRTBP. Middle panel, the surface of
    section for the same initial conditions computed with the normal
    form approximating the CPRTBP Hamiltonian. Right panel, the
    comparison between the two surfaces of section.  In the three
    plots, $L_4$ is located at (${4\pi}/{3}$, $0$).}
  \label{fig:surf_sec}
\end{figure}

In Fig.~\ref{fig:surf_sec} we show the results of the comparison
between the complete CPRTBP Hamiltonian and the normal form
approximating it, in the case of the Earth-Moon system, which is
defined by its value of the mass parameter $\mu =
0.01215058561$.  In the left panel, we show the surface of section
numerically computed by considering the equations of motion related to
the CPRTBP Hamiltonian. We take a set of 10 equispaced initial
conditions, with $\rho=0$, $4.188 \leq \lambda \leq 4.45$, $\ell =
0$. The value for $\Gamma$ has been set in such a way that all the
initial conditions keep the same value for the Jacobi constant as in
$L_4\,$. These orbits has been integrated up to recover 1000 points
over the surface defined by $\ell = 0$. Those points are drawn in
black in the space of variables ($\lambda$,$\rho$).

For the same initial conditions as before, we have integrated the
orbits also according to the semi-analytical
scheme~\eqref{semi-analytical_scheme}, up to collect also 1000 points
over the surface. This computations have been automatically made by a
code written in {\bf C}, in such a way to use the expansions of the
normal form $\Zscr^{(R_1,R_2)}$ and the canonical transformation
$\Cscr^{(R_1,R_2)}$, which were preliminarly produced by using {\tt
  Mathematica}. The truncations have been made according to the
following values of the parameters ruling the extension of those
expansions: $R_1 = 3$, $R_2 = 5$, $K = 5$, $l_{max} = 5$ and $s_{max}
= 3$. These values imply reasonable computing times. The results are
expressed in the middle panel of Fig.~\ref{fig:surf_sec}. Finally, in
the right panel, we show the comparison between the two surfaces of
section. For all the orbits close to the equilibrium point $L_4\,$,
the correlation is extremely good, thought the approximation fails on
reproducing exactly the tadpole orbits far from $L_4\,$.

\section{Optimal transfers by means of integrable aproximations}\label{sec:opt_trans}

\subsection{Baseline}\label{sbs:vcht_alg}

Since the new normal form has $1$ d.o.f., it provides the integrable
approximations of the motions of small bodies in the vicinity of the
stable equilibrium points and suitable surfaces of section that are
easy to compute. These tools can be used in many different studies. As
explained in the Introduction, we focus in the design of maneuvres of
small bodies (as spacecrafts or space debris) that are originally out
of the stability region filled by tadpole orbits and we would like to
situate into it. With such an aim, we design a method that allows to
study the effectiveness of a set of impulses, done after considering a
fixed initial position.

As outline of the algorithm, we have to
\begin{itemize}
\item[a.] choose a starting point $P_s$, which corresponds to the
  position of the body to transfer at $t=0$ and translate it to
  cartesian coordinates $P_s=\big(x_0,y_0,V_{x_0},V_{y_0}\big)$.
\item[b.] give an impulse
\begin{equation} 
\Delta V_{x_0} = \|\Delta\vet{V}\|\cos \beta
\qquad
\Delta V_{y_0} = \|\Delta\vet{V}\|\sin \beta \ , 
\label{eq:impulse}
\end{equation}
where $\|\Delta\vet{V}\|$ and $\beta$ are the modulus and direction of
the impulse, respectively, in cartesian coordinates.  This gives a new
orbit, whose evolution in normalized variables can be computed with
the integrable approximation of the Hamiltonian and checked in the
surfaces of section.
\item[c.] decide the acceptance or rejection of the impulse, according
to whether it fulfills or not the conditions of the transfer.

\end{itemize}

As a matter of fact, this mechanism can be applied for a physically
adequate range of moduli and directions and not one orbit by one, in order
to speed up the computations.\\

\subsection{Application and results}\label{sbs:results}

\begin{figure}
  \includegraphics[height=.3\textheight,angle=270]{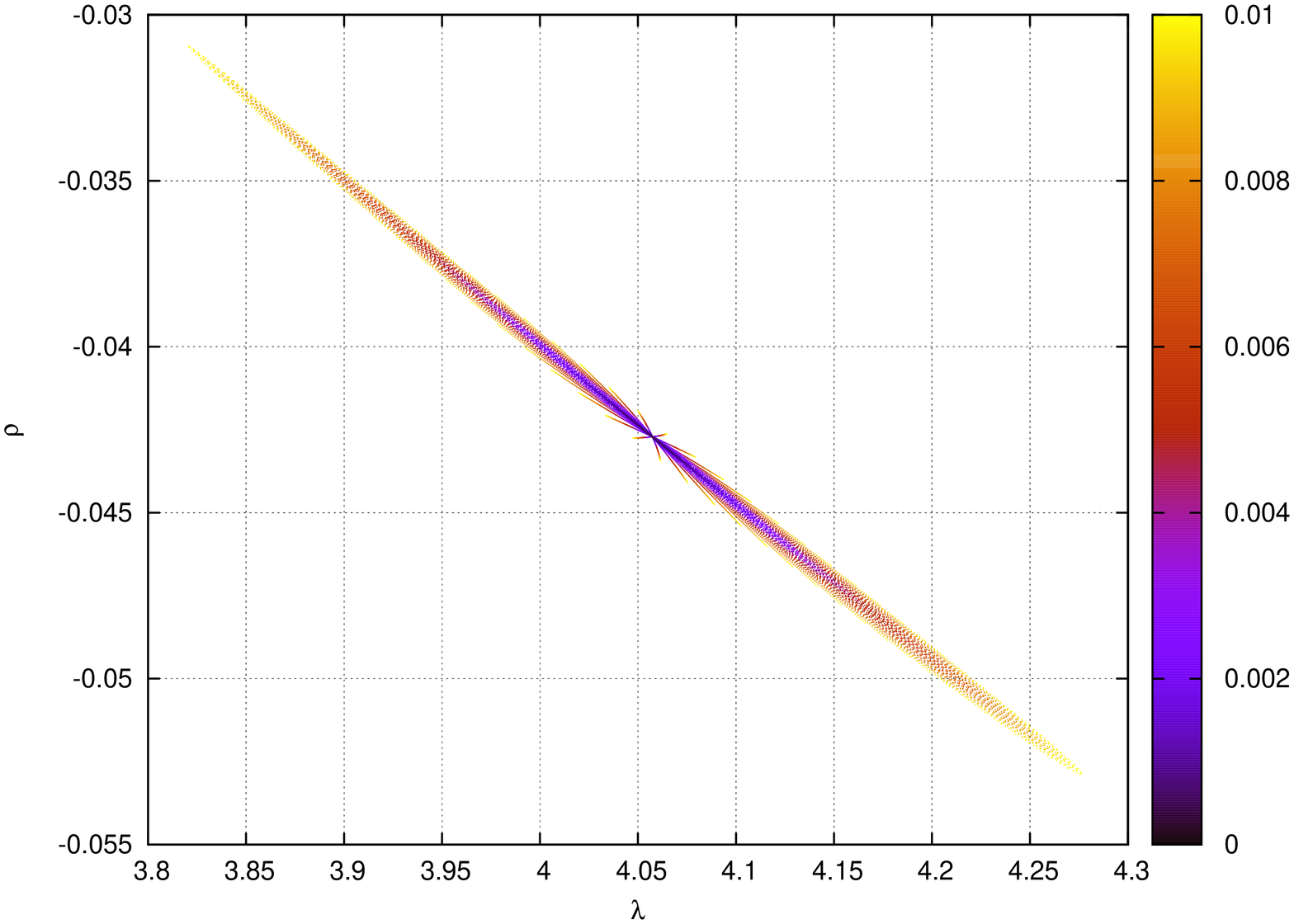}
  \includegraphics[height=.3\textheight,angle=270]{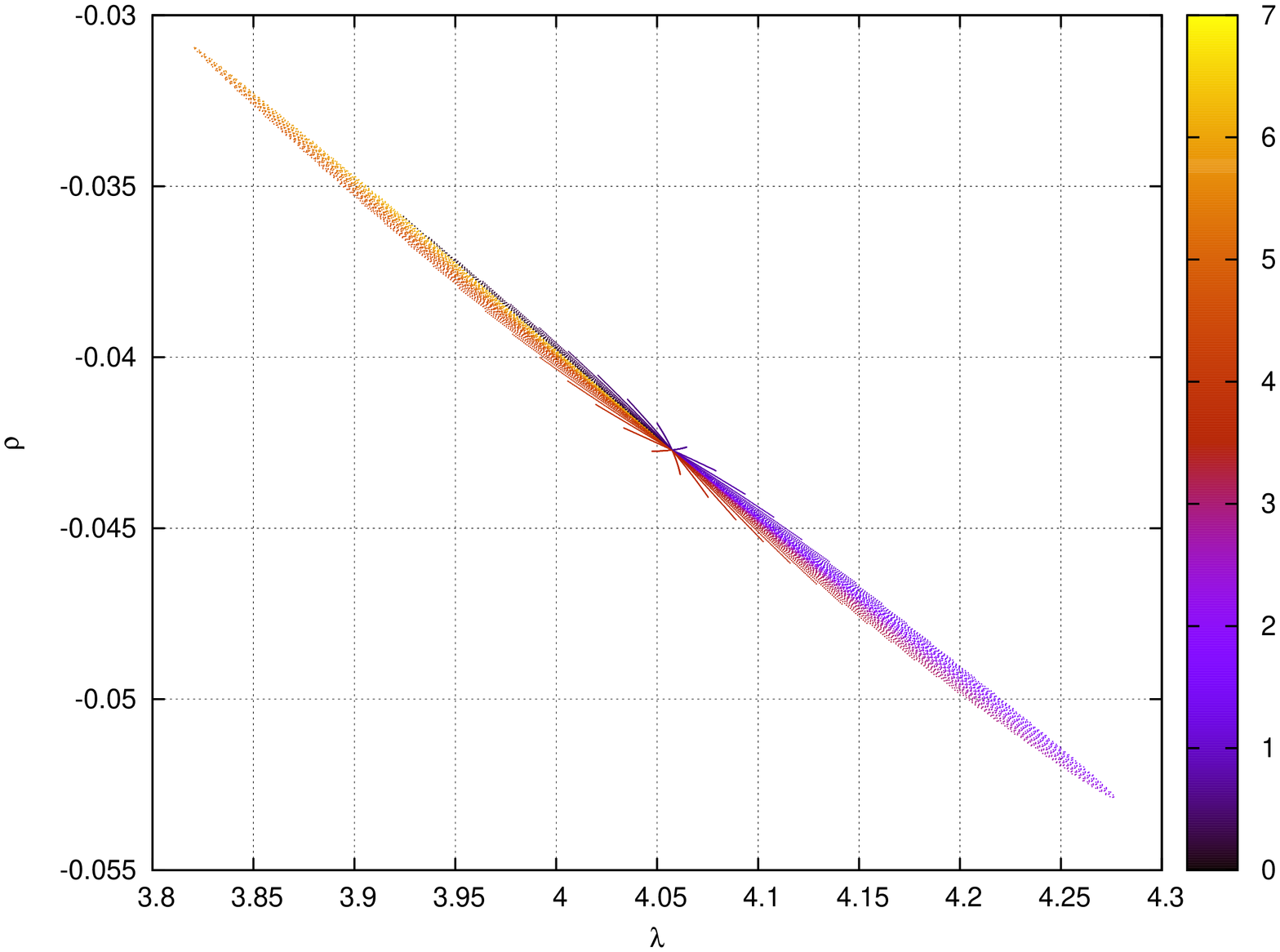}
  \caption{Color scheme for the initial conditions produced after
    applying the impulses. Left panel, new positions in
    ($\lambda$,$\rho$). The color scale represents the magnitude of
    the impulse $\|\Delta\vet{V}\|$. Right panel, same as before with
    color scale representing the direction of the impulse $\beta$.}
  \label{fig:impulses}
\end{figure}

We apply the method described above to the case of the Earth-Moon
system. As starting point $P_s\,$, we choose a position slightly
outside the real stability region\footnote{More specifically, $P_s$ is
  in the chaotic region related to the stable/unstable manifolds
  emanating from $L_3$. It has been fixed in order to be one of the
  points closest to $L_4$ among those belonging to the chaotic set
  and to the axis of the abscissas of the surface of section defined
  by $\ell = 0$.}, estimated by numerical integrations of the full
problem. $P_s$ is given by $\lambda = 3.95$, $\rho = 0$, $\csi =
0.0885442$ and $\eta = 0$.  For the transfer orbits, we apply $10^4$
different impulses, for which $0 \leq
\|\Delta\vet{V}\| \leq 0.01\,$ and $0 \leq \beta < 2\pi\,$.  Following
\S2.8 in \cite{MurrDermm-1999}, \eqref{eq:Delaunay-coord}
and~\eqref{eq:def-total-canonical-transf}, we translate every orbit to
normalized variables. In Fig. \ref{fig:impulses}, we present the
results of the translation of the new coordinates calculated after
each impulse (for the first new point of the orbit on the surface of
section) in the corresponding variables. The color scale represents
the size of the impulse $\|\Delta\vet{V}\|$ in the left panel and its
direction $\beta$ in the right panel. As expected, a stretching effect
occurs while translating to the new system, and the impulses are more
effective (i.e., same values imply a bigger difference with respect to
the original point $P_s$) in some directions.

\begin{figure}
  \includegraphics[height=.3\textheight,angle=270]{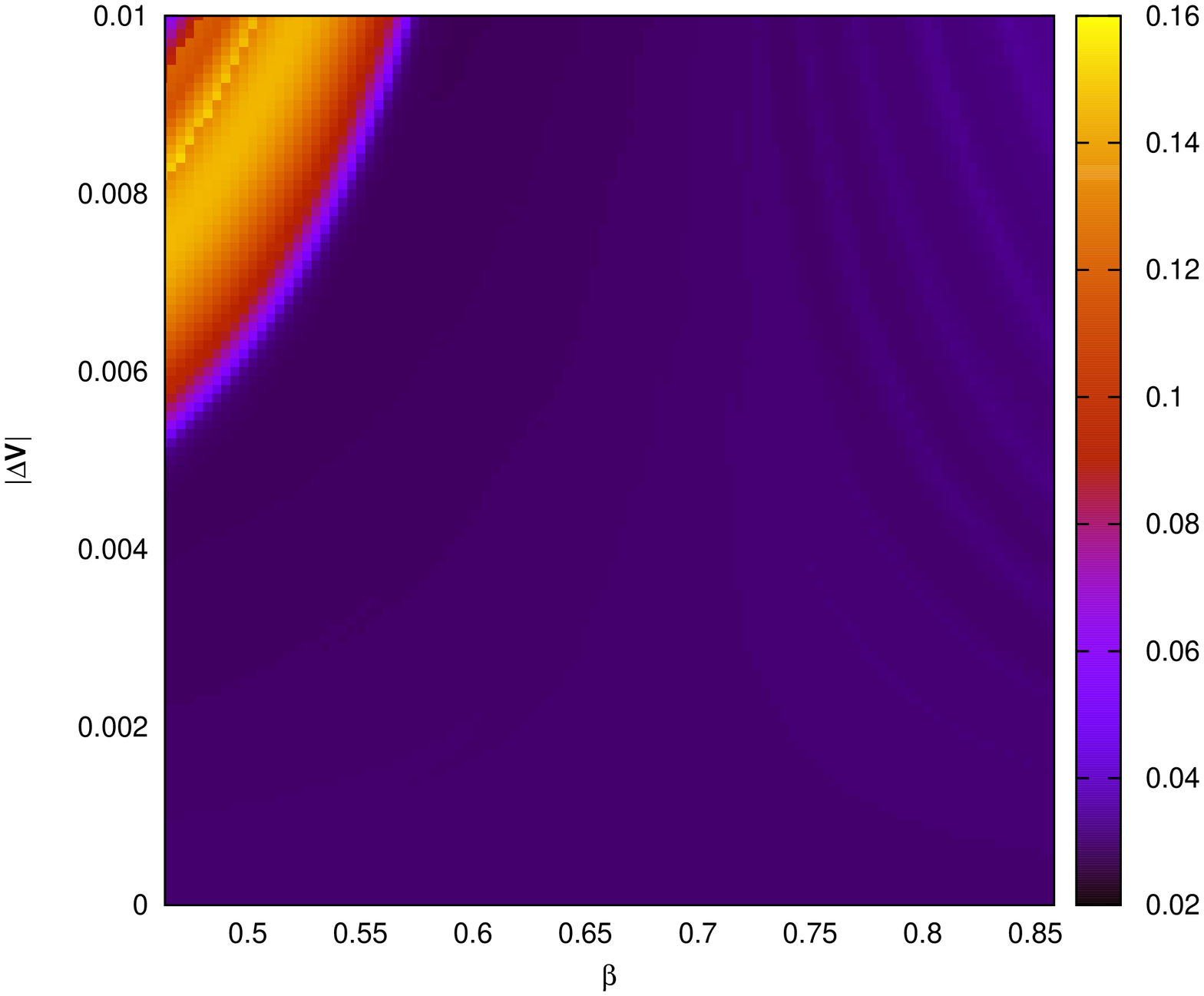}
  \includegraphics[height=.3\textheight,angle=270]{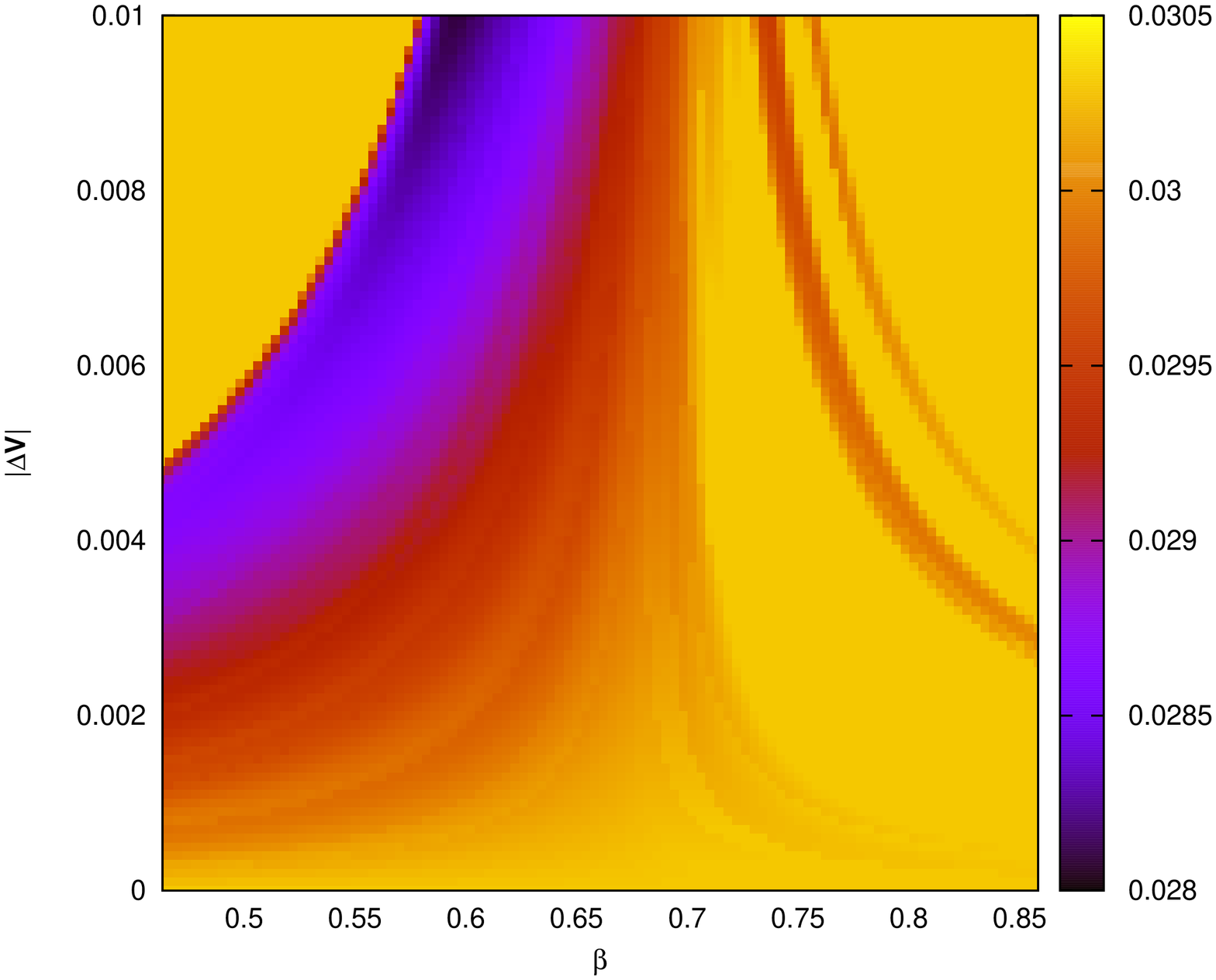}
  \caption{Left panel, values of the areas enclosed in the surface of
    section for every considered combination of $\|\Delta\vet{V}\|$
    and $\beta$, the color scale represents the results of the
    calculations using the averaged normal form. Right panel, same as
    before where the computed areas are reported only if they are
    smaller than the value ($A_0$) corresponding to the original
    initial condition $P_s\,$.}
  \label{fig:areas}
\end{figure}

In order to distinguish which impulses generate orbits that go deeper
inside the stability region, we isolate the impulses pointing towards
the inner part. We refine the values of the angles considered, taking
$100$ equidistant values $\beta$, where $0.46 \leq \beta \leq
0.86\,$. We integrate their evolution with respect to the normal form
$\Zscr^{(5,3)}$ and we find the curves described by these orbits on the
surface of section.  In the averaged normal form, each of these curves
depicts the area associated to one of the actions defining the 2-D
torus that is invariant with respect to the motion.  Since as closer
the torus is to the equilibrium point $L_4\,$, as smaller this area
is, we take as criterion for a good transfer a reduction of this value
with respect to the initial one (\emph{Minimum
  Action criterion}, since such an area corresponds to an action).  In
Fig.~\ref{fig:areas}, we show the results of the computation of the
area in two different color-scales. In the left panel, we can see for
all the considered combinations of modulus $\|\Delta\vet{V}\|$ and
angles $\beta$, the value of the area described by the transfered
orbit. The lower border of the plot represents the orbits with
$\|\Delta\vet{V}\| = 0$, so their areas ($A_0$) are equal and can
be used as reference value. In the right panel, only orbits
corresponding to areas $\leq A_0$ are considered, the greater ones are
fixed equal to $A_0\,$. This allows a better discrimination between
all the orbits providing suitable impulses.

Both plots shows that the selection of the angle for the impulse is
not trivial, since small differences generate very different results.
Consider, for instance, the impulses with $\beta=0.50$ and with
$\beta=0.55\,$, for $\|\Delta\vet{V}\| > 0.006$). The former give the
best approach to the optimal transfer we look for, while the latter
generate orbits which are actually further from $L_4$ than that
related to the initial condition $P_s\,$. In general, since we take
small values for the impulses, in the correct directions, bigger
$\|\Delta\vet{V}\|$ implies smaller final areas. The best choices for
the transfer correspond to the impulses for which
$\beta\in[0.50\,,\,0.68]\,$, and suitable values for their sizes,
following the darker stripes in Fig.~\ref{fig:areas}.

\section{Conclusions}\label{sec:conclus}

In this work we explicitly construct an integrable normal form
approximating the CPRTBP Hamiltonian. This is done by reformulating
the approach described in~\cite{Garfinkel-77} so as to use some more
modern techniques, developed in the framework of the
Hamiltonian perturbation theory and mainly based on the Lie series
formalism. This allow us to design an algorithm that can be fully
translated in programming codes. In particular, we 
produce a truncation of the normal form $\Zscr^{(5,3)}$, whose the
expansion in~\eqref{eq:split-integr-and-perturb-parts-in-H(r1,r2)}
highlights that it is an average of the CPRTBP Hamiltonian with
respect to the angle associated to its fast dynamics. The first
results provided by this revisited approach are encouraging: in some
suitable surface of sections our algorithm provides very good
approximations of the tadpole orbits close enough to $L_4-L_5\,$.
However, we are aware of the fact that the accuracy of our expansions
must be strongly improved in order to face challenging concrete
problems in a region far from those equilibrium points. In our
opinion, the main constraint on the quality of our results is due to
the truncations on the Fourier series in the slow angle $\lambda$. We
think that this limitation can be removed, by representing the
dependence on $\lambda$ in a suitable way, so as to avoid Fourier
expansions. We plan to investigate such a new approach in the near
future.

Furthermore, we show a first astrodynamical application starting
from our calculation of the integrable normal form $\Zscr^{(5,3)}$,
which approximates the CPRTBP Hamiltonian.  We design an algorithm
that allowed to compute optimal transfers between orbits in the
neighborhood of the equilateral Lagrangian equilibrium points. We
generate impulses in cartesian variables according to
formula~\eqref{eq:impulse}, on a grid of values for
$\|\Delta\vet{V}\|$, the magnitude of the impulse, and $\beta$, the
direction. For those, we are able to discriminate the
\emph{suitable transfers}, that imply a final orbit closer to the
equilibrium point, according to a new criterion. Using our normal form
as an approximation of the complete CPRTBP, we estimate the area
enclosed by the final orbit, and minimizing this quantity, we
select the best candidates for the transfer. A careful inspection of
the plots shows that the best candidates are highly depending on the
size of the impulse, and very sensitive to the changes on the angle
$\beta$.


\section*{Acknowledgments}
The authors would like to thank C.~Efthymiopoulos, because of his
constant support which allowed us to finalize our {\tt Mathematica}
codes. We are indebted also with C.~Sim\`o who suggested to one of us
to reconsider the fundamental work~\cite{Garfinkel-77} by
Garfinkel. During this work, R.I.P. was supported by the Astronet-II
Marie Curie Training Network (PITN-GA-2011-289240), while U.L. was
partially supported also by the research program ``Teorie geometriche
e analitiche dei sistemi Hamiltoniani in dimensioni finite e
infinite'', PRIN 2010JJ4KPA\_009, financed by MIUR.






\end{document}